\documentclass[trackchanges,twocolumn]{aastex7}
\usepackage{lipsum}
\usepackage{amsmath}
\usepackage{enumitem}

\begin{document}

\title{Dependency of the Bar Formation Timescale On The Halo Spin}


\author[0000-0001-8962-663X]{Bin-Hui Chen}
\affiliation{Tsung-Dao Lee Institute, Shanghai Jiao Tong University, Shanghai 200240, People’s Republic of China}
\affiliation{Department of Astronomy, School of Physics and Astronomy, Shanghai Jiao Tong University, 800 Dongchuan Road, Shanghai 200240, People’s Republic of China}
\affiliation{State Key Laboratory of Dark Matter Physics, School of Physics and Astronomy, Shanghai Jiao Tong University, Shanghai 200240, People's Republic of China}
\affiliation{Key Laboratory for Particle Astrophysics and Cosmology (MOE) / Shanghai Key Laboratory for Particle Physics and Cosmology, Shanghai 200240, People's Republic of China}
\email{2000cbh@sjtu.edu.cn}

\author[0000-0003-3657-0200]{Sandeep Kumar Kataria}
\affiliation{Department of Space, Planetary \& Astronomical Sciences and Engineering, Indian Institute of Technology Kanpur, Kanpur 208016, India}
\email[show]{skkataria.iit@gmail.com}

\author[0000-0001-5604-1643]{Juntai Shen}
\affiliation{Department of Astronomy, School of Physics and Astronomy, Shanghai Jiao Tong University, 800 Dongchuan Road, Shanghai 200240, People’s Republic of China}
\affiliation{State Key Laboratory of Dark Matter Physics, School of Physics and Astronomy, Shanghai Jiao Tong University, Shanghai 200240, People's Republic of China}
\affiliation{Key Laboratory for Particle Astrophysics and Cosmology (MOE) / Shanghai Key Laboratory for Particle Physics and Cosmology, Shanghai 200240, People's Republic of China}
\email[show]{jtshen@sjtu.edu.cn}

\author{Meng Guo}
\affiliation{Shandong Computer Science Center (National Supercomputing Center in Jinan), Qilu University of Technology (Shandong Academy of Sciences), Jinan, Shandong 250013, People’s Republic of China}
\affiliation{Jinan Institute of Supercomputing Technology, Jinan, Shandong 250103, People’s Republic of China}
\email{guomeng@sdas.org}

\correspondingauthors{Sandeep Kumar Kataria, Juntai Shen}

\def\fdisk{f_\mathrm{disk}}
\def\tbar{\tau_\mathrm{bar}}
\def\thickness{h_z/R_\mathrm{d}}

\def\spinMainValue{0.1}

\def\toleranceOfTauBarErr{13.7\ \mathrm{Gyr}}
\def\tauBarMax{100\ \mathrm{Gyr}}
\def\Sbar0Max{0.1}
\def\remainNum{531}

\def\Afit{1216}
\def\Aerr{157.96}
\def\Bfit{-9.07}
\def\Berr{0.35}

\def\multiSpinModel{\mathrm{fd0.60Q1.6}}

\begin{abstract}

Bars are among the most prominent structures in disk galaxies. While the widely accepted swing-amplification theory provides a qualitative framework for their formation, the detailed physical processes remain incompletely understood. Previous studies have shown that the bar formation timescale in isolated galaxies depends exponentially on the disk mass fraction (the so-called ``Fujii relation") and linearly on disk hotness and thickness. However, the influence of dark matter halo spin on bar formation has not been systematically investigated. In this work, we construct a suite of \textit{N}-body disk–halo models with varying disk mass fractions and amounts of random motions. By introducing prograde and retrograde spins in the dark matter halo, we explore how halo spin modifies the established empirical relations governing bar formation timescales. We find that these relations remain valid in both prograde and retrograde halo spin models. For rapid bar formation (short timescale), the effect of halo spin is nearly negligible. In contrast, for moderately slow bar formation, prograde (retrograde) halo spin tends to accelerate (suppress) bar onset. In cases of extremely slow bar formation, halo spin introduces a stronger but more stochastic influence. These trends might arise from the exchange of angular momentum between the stellar disk and the dark matter halo.

\end{abstract}


\section{Introduction} \label{sec:intro}

As one of the most prominent galactic features, bars are found in approximately two-thirds of disk galaxies \citep[e.g.,][]{eskrid_etal_2000, menend_etal_2007, barazz_etal_2008, sheth_etal_2012, simmon_etal_2014, erwin_2018, lee_etal_2019}. They are present across cosmic time, from the local Universe to high redshifts, as demonstrated by recent JWST observations that reveal numerous barred galaxies at early epochs \citep{guo_etal_2023, lecont_etal_2024}. Bars play a critical role in the secular evolution of their host galaxies \citep[e.g.,][]{hohl_1971, lyn_kal_1972, combes_etal_1990, kor_ken_2004, athana_2005, portai_etal_2015, fragko_etal_2019, gadott_etal_2020, sai_cat_2022, neuman_etal_2024}, including triggering Lindblad resonances, redistributing stellar mass \citep{debatt_etal_2017, berald_etal_2023}, funneling gas into galactic centers to feed supermassive black holes \citep{athana_1992, li_etal_2016, li_etal_2017, li_etal_2023, kat_viv_2024}, heating the stellar disk \citep{min_fam_2010, kim_etal_2024}, and shaping complex stellar populations \citep{debatt_etal_2020, che_li_2022, chen_etal_2024}. A comprehensive understanding of bar formation and evolution is thus essential for unraveling the broader picture of galaxy evolution.

Among the various properties of galactic bars, their formation timescale is critical. It is well established that bars can naturally emerge from dynamical instabilities in cold, rotationally supported stellar disks \citep[e.g.,][]{com_san_1981, efstat_etal_1982, raha_etal_1991, shen_etal_2010}. Following their formation, bars often undergo vertical thickening \citep{sel_ger_2020}, giving rise to boxy, peanut-shaped, or ``X"-shaped bulges \citep{combes_etal_1990, li_she_2012, li_she_2015, Kumar.et.al.2022, Kataria.2024}. After buckling, the bar typically enters a phase of steady secular evolution. The timing and conditions under which a bar forms, therefore, have significant implications for the host galaxy’s evolution. For instance, once a bar has fully developed, it can drive large-scale gas inflows that fuel central active galactic nuclei (\citealt{athana_1992, li_etal_2016, li_etal_2017, li_etal_2023, kat_viv_2024}; may incorporate with other mechanisms, see \citealt{lima_etal_2022}), potentially triggering stellar feedback and quenching star formation across the galaxy.

Previous studies have established several empirical relations for the bar formation timescale. \cite{fujii_etal_2018} found that in \textit{N}-body simulations of disk-dark matter (DM) halo systems, the bar formation timescale is tightly correlated with the disk mass fraction
\begin{equation}
\fdisk \equiv \dfrac{V_{\mathrm{c,disk}}^2}{V_{\mathrm{c,tot}}^2}\bigg|_{R=2.2R_{\mathrm{d}}}
\end{equation} where $R_\mathrm{d}$ is the exponential scale length of the disk. Defining the bar formation timescale $t_\mathrm{bar}$ as the first epoch when the radially maximal value of the bar strength
\begin{equation}
A_2\equiv\left\vert\dfrac{\sum_j m_j\exp(2i\phi_j)}{\sum_j m_j}\right\vert
\end{equation}
exceeds 0.2, they found that $t_\mathrm{bar}$ depends exponentially on the disk mass fraction as
\begin{equation}\label{eq: Fujii formula}
t_{\mathrm{bar}}/\mathrm{Gyr}=(0.146 \pm 0.079)\exp \left[(1.38 \pm 0.17) / \fdisk\right].
\end{equation} \cite{bland_etal_2023} later proposed a more physically motivated definition of the bar formation timescale, $\tbar$, based on the exponential growth phase of $A_2$ predicted by classical swing amplification theory \citep{binney_2020, bland_etal_2023}:
\begin{equation}\label{eq: definition of tau bar}
A_2(t) = A_2(0)\cdot\displaystyle{\exp(\frac{t}{\tbar})}.
\end{equation} They confirmed the exponential dependence of bar formation on $\fdisk$ and referred to this correlation as the ``Fujii relation.'' Additionally, they proposed a modified version better suited to their ``high-mass" models:
\begin{equation}\label{eq: BH formula}
\tbar= \tau_{\mathrm{H}}\cdot\exp \left[-13.0\left(f_{\text {disk }}- 0.328\right)\right],
\end{equation}
where $\tau_\mathrm{H}$ is the Hubble timescale. The Auriga simulations \citep{grand_etal_2017, grand_etal_2024} exhibit a similar decreasing trend in bar formation timescale with increasing disk mass fraction \citep{fragko_etal_2025}. Simulations and observations also indicate that bar formation is slowed in dynamically hotter disks (i.e., with higher Toomre $Q$ values, \citealt{ath_sel_1986, sheth_etal_2012, worrak_2025}). \cite{che_she_2025} extended these results by systematically investigating the dependence of $\tbar$ on the Toomre $Q$ parameter and the disk scale height $h_z$. In a wide parameter range ($0.8 \lesssim Q \lesssim 2.0$ and $0.2 \lesssim h_z \lesssim 1.0$ for disks with $R_\mathrm{d} = 3.0\ \mathrm{kpc}$), they confirmed that the Fujii relation remains valid. Additionally, they found that $\tbar$ scales approximately linearly with both $Q$ and $h_z$. Combining these dependencies, \cite{che_she_2025} proposed the following empirical relation:
\begin{equation}
\label{eq: fit in CS2025}
\dfrac{\tbar}{\Afit\ \mathrm{Gyr}} = Q\dfrac{h_z}{R_\mathrm{d}}\exp(-\dfrac{\fdisk}{0.11}),
\end{equation}

While these studies have revealed insights into the physics of bar formation, they do not provide a complete picture. Bar formation is a complex process influenced by numerous additional factors, including the central mass concentration of the stellar disk \citep[e.g.][]{kat_das_2018, kat_das_2019, katari_etal_2020, jan_kim_2023} and the gas mass fraction of the galaxy \citep{bland_etal_2023, bland_etal_2024, bland_etal_2025}. Among the key factors not explored in \cite{che_she_2025}, the spin of the DM halo is particularly important \citep{sah_naa_2013, kat_she_2022, romeo_etal_2023, Ansar.et.al.2023, Ansar.Das.2024, chi_kat_2024}. The halo spin is commonly quantified by the dimensionless parameter
\begin{equation}
\lambda = \dfrac{J}{\sqrt{2GMR}},
\end{equation} where $J$, $M$, and $R$ are the DM halo's specific angular momentum, virial mass, and virial radius, respectively. Halo spin originates from tidal torques during the assembly history of galaxies \citep{hoyle_1949, sciama_etal_1955, peeble_1969, dorosh_1970, white_1984, bar_efs_1987, maller_etal_2002, vitvit_etal_2002, sch_bjo_2009}. The spin parameter follows a lognormal distribution, peaking at $\lambda \sim 0.035$ \citep{bulloc_etal_2001, het_bur_2006, bett_etal_2007, kne_pow_2008, ishiya_etal_2013, zju_spr_2017}. The development of a bar is intimately linked to the transfer of angular momentum from the inner disk to the outer disk and DM halo \citep{athana_2002, athana_2003_a, athana_2003_b, long_etal_2014}. Thus, as a proxy for the halo's angular momentum content, $\lambda$ significantly influences bar formation. Several studies have shown that prograde (positive) halo spin tends to promote bar formation, whereas retrograde (negative) spin slows it \citep{sah_naa_2013, kat_she_2022}. Given these findings, it is natural to ask how halo spin modulates the empirical bar formation timescale relations proposed by \citet{fujii_etal_2018}, \citet{bland_etal_2023}, and \cite{che_she_2025}.

In this work, we investigate how halo spin influences the empirical relations governing bar formation timescales. Building on several models from \cite{che_she_2025}, we introduce halo spin by reversing the angular momentum of selected inner DM particles, following the method of \citet{kat_she_2022, kat_she_2024}. We then examine how halo spin modifies the established empirical relations, and explore how it shapes $\tbar$ by regulating the angular momentum transfer from the disk to the halo.

The structure of this paper is as follows. In Section~\ref{sec: model}, we briefly describe the setup of the \textit{N}-body models. Section~\ref{sec: results} presents the main results and their implications. We summarize this paper in Section~\ref{sec: summary}.

\begin{table}
\caption{Main parameters of the \textit{N}-body models}\label{table: parameters}
\centering
\begin{tabular}{ccccc}
\hline
Component & $N_\mathrm{p}$ & $M/10^{10} M_\odot$ & $a/\mathrm{kpc}$ & $r_\mathrm{c}/\mathrm{kpc}$
\\
DM halo & $500,000$ & 50.0 & 20.0 & 250.0
\\
\hline
Component & $N_\mathrm{p}$ & $R_\mathrm{d}/\mathrm{kpc}$ & $h_z/\mathrm{kpc}$ 
\\
Stellar disk & $500,000$ & 3.0 & 0.4
\\
\hline
\end{tabular}
\end{table}

\section{Model Setup}\label{sec: model}

The models used in this study are based on those presented in \cite{che_she_2025}. Each system consists of a DM halo following a Navarro–Frenk–White (NFW) profile \citep{navarr_etal_1996} and a quasi-isothermal stellar disk \citep{vasili_2019}, with the key structural parameters listed in Table~\ref{table: parameters}. Further details of the model construction can be found in \cite{che_she_2025}.

Following \citet{kat_she_2022}, \citet{chi_kat_2024}, and \citet{kat_she_2024}, we spin up 60 \textit{N}-body models taken from \cite{che_she_2025}, covering $\fdisk = 0.35,\ 0.40,\ \ldots,\ 0.80$ (10 values) and $Q = 1.0,\ \ldots,\ 2.0$ (6 values), with $h_z \equiv 0.4\ \mathrm{kpc}$. Note that these are the initial parameters, and the actual values of $\fdisk$, $Q$, and $h_z$ may slightly deviate from the specified targets, remaining within predefined tolerances. Compared to \cite{che_she_2025}, we adopt tighter tolerances between the target values and the actual ones: $\epsilon(\fdisk) = 0.025$, $\epsilon(Q) = 0.05$, and $\epsilon(h_z) = 0.025\ \mathrm{kpc}$, to minimize random fluctuations in $\tbar$ due to small deviations in $\fdisk$, $Q$, and $h_z$. Each model is named as $\mathrm{fdX_1QX_2}$ with $X_1=\fdisk$ and $X_2=Q$ (for simplicity, referring to the target values). For each model, the halo is spun to $\lambda = \pm\spinMainValue$, with positive values corresponding to prograde spin and negative values to retrograde spin.

\begin{figure*}
\centering
\includegraphics[height=.99\textheight]{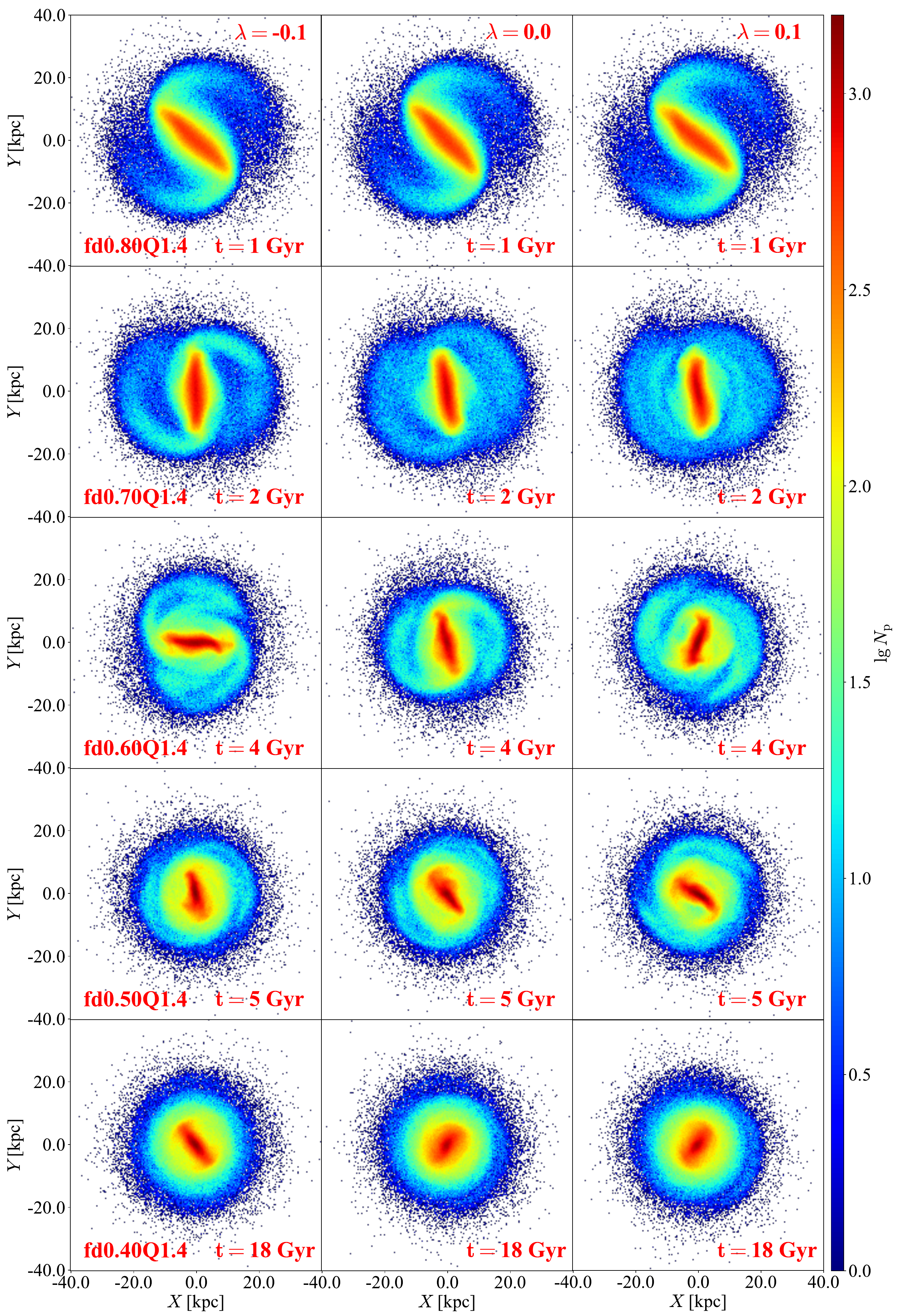}
\caption{Face-on views of the forming bars near the saturation of their exponential growth for several models with $Q = 1.4$. From bottom to top, $\fdisk$ increases. From left to right: models with retrograde, zero, and prograde halo spin.}
\label{fig: face-on views}
\end{figure*}

We evolve the \textit{N}-body models using the \texttt{GADGET4} code \citep{spring_etal_2021}. Each model is integrated over an adaptive time interval ranging from $8$ to $40\ \mathrm{Gyr}$ to ensure sufficient time for bar formation. Figure~\ref{fig: face-on views} presents face-on views of stellar bars in several models with $Q = 1.4$, captured near the saturation phase of their exponential growth at different times.

\begin{figure*}
\centering
\includegraphics[width=.95\textwidth]{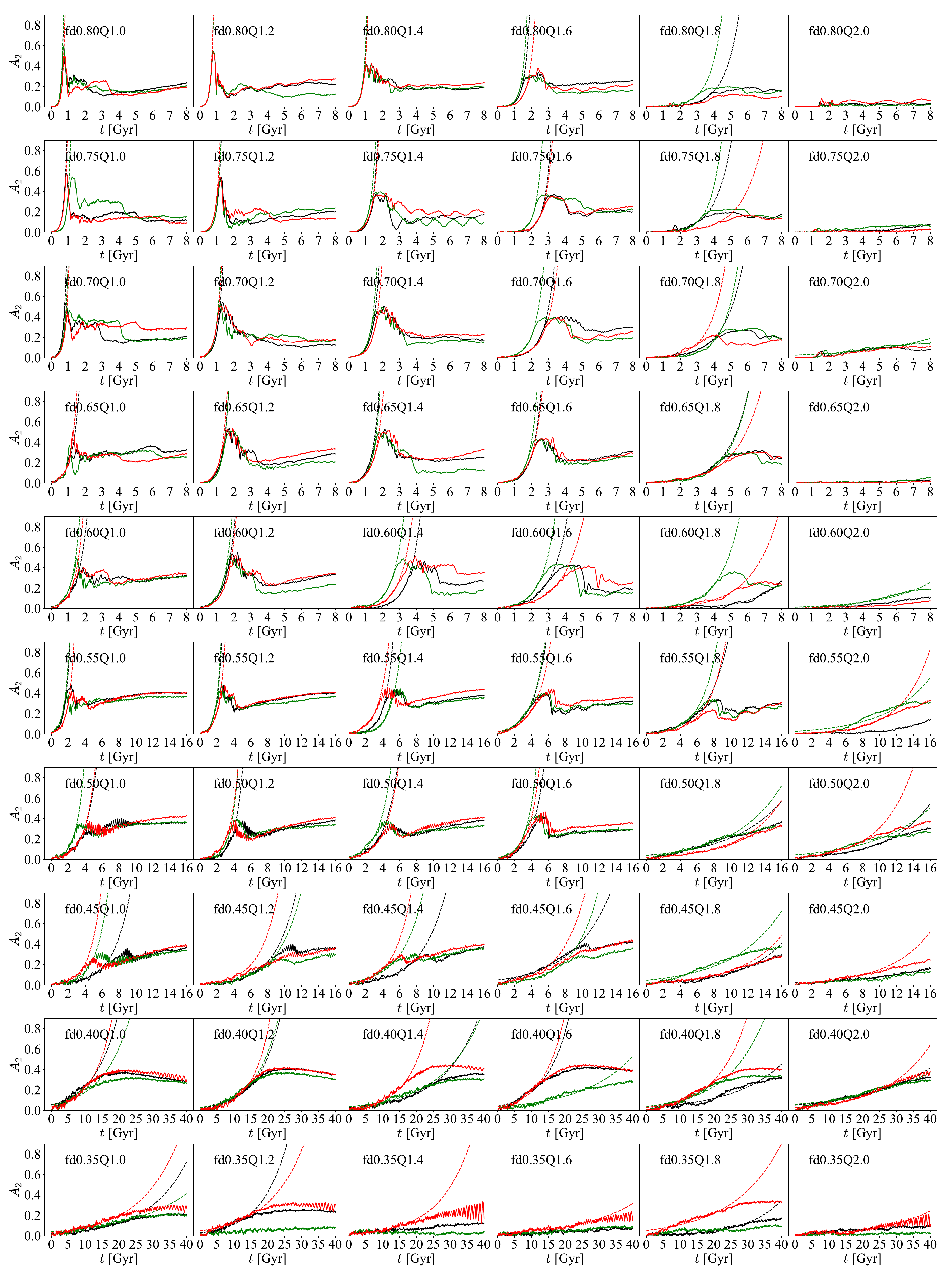}
\caption{Time evolution of bar strength $A_2(t)$ for models with different halo spins: $\lambda = 0$ (black), $\lambda = 0.10$ (green), and $\lambda = -0.10$ (red). From top to bottom, panels are ordered by decreasing $\fdisk$; from left to right, by increasing Toomre $Q$. For models exhibiting effective bar formation ($\operatorname{max}\{A_2(t)\} \geq 0.15$), dashed lines indicate exponential fits to the initial growth phase of $A_2(t)$ in the corresponding color.}
\label{fig: spin effect}
\end{figure*}

\begin{figure}
\centering
\includegraphics[width=.495\textwidth]{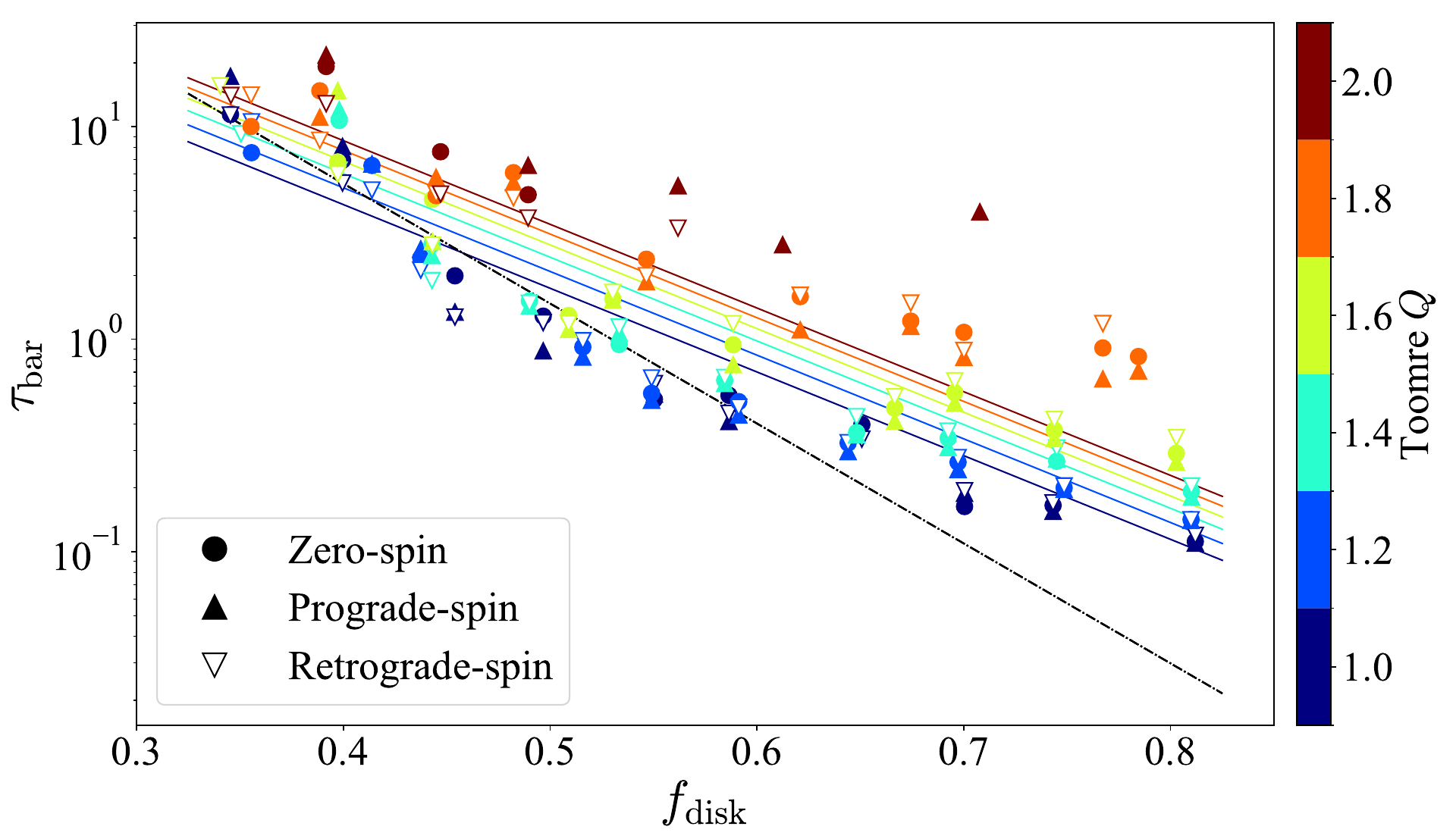}
\caption{The ``Fujii diagram": distribution of $\tbar$ versus $\fdisk$ for models with effective bar formation. Data point colors indicate Toomre $Q$ values. Marker shapes denote halo spin: circles for $\lambda = 0$, upward-filled triangles for $\lambda = 0.10$, and downward-unfilled triangles for $\lambda = -0.10$. For comparison, we show the Fujii relation from the high halo mass models of \citet{bland_etal_2023} in a black dotted-dashed line (Equation~\ref{eq: BH formula}) and the empirical relation from \cite{che_she_2025} in colored solid lines (Equation~\ref{eq: fit in CS2025}). Overall, models with different spin values follow the general empirical trends, with halo spin introducing additional variations in $\tbar$.}
\label{fig: Fujii diagram}
\end{figure}

\begin{figure*}
\centering
\includegraphics[width=.48\textwidth, height=.22\textheight]{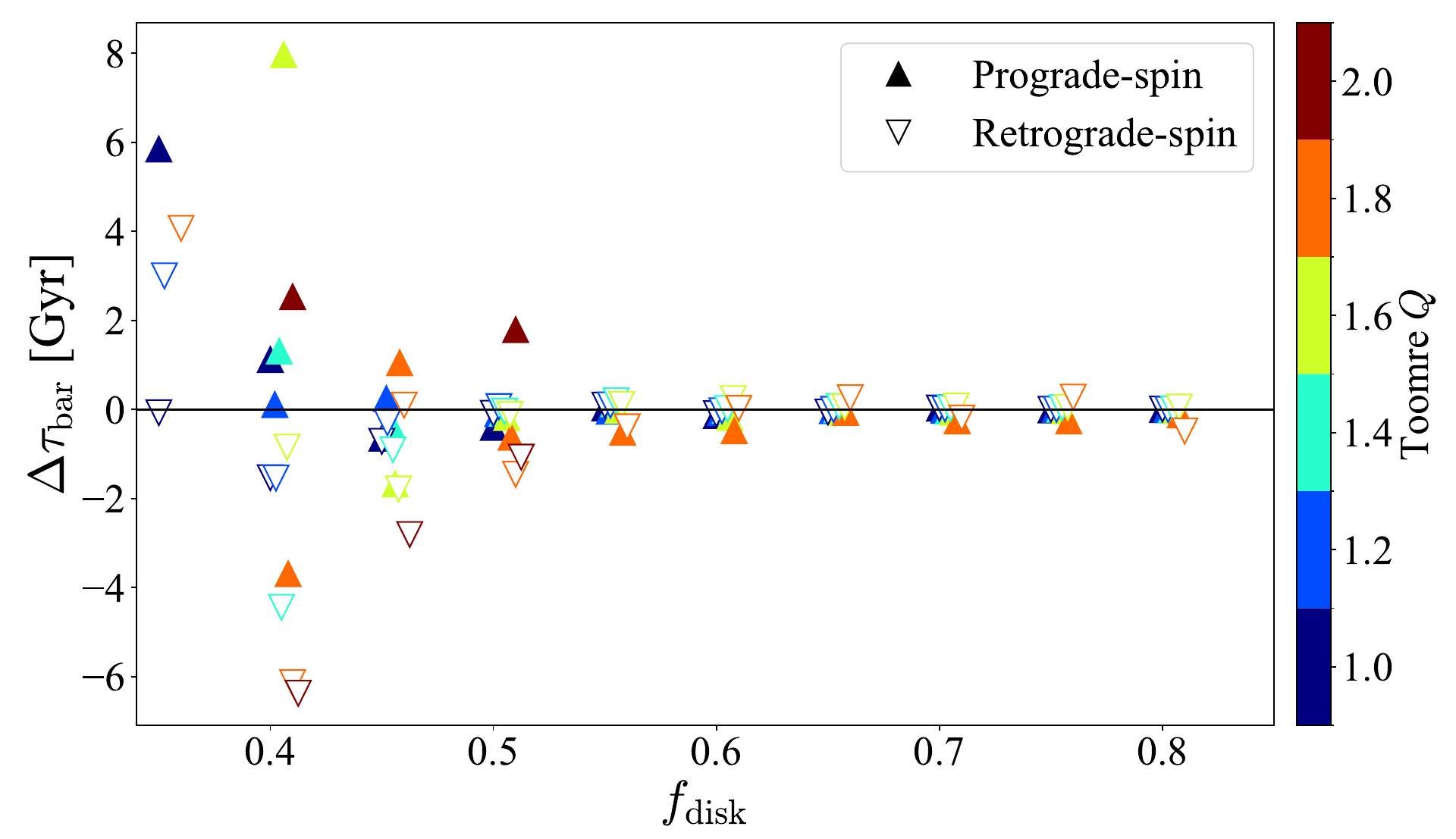}
\includegraphics[width=.48\textwidth, height=.22\textheight]{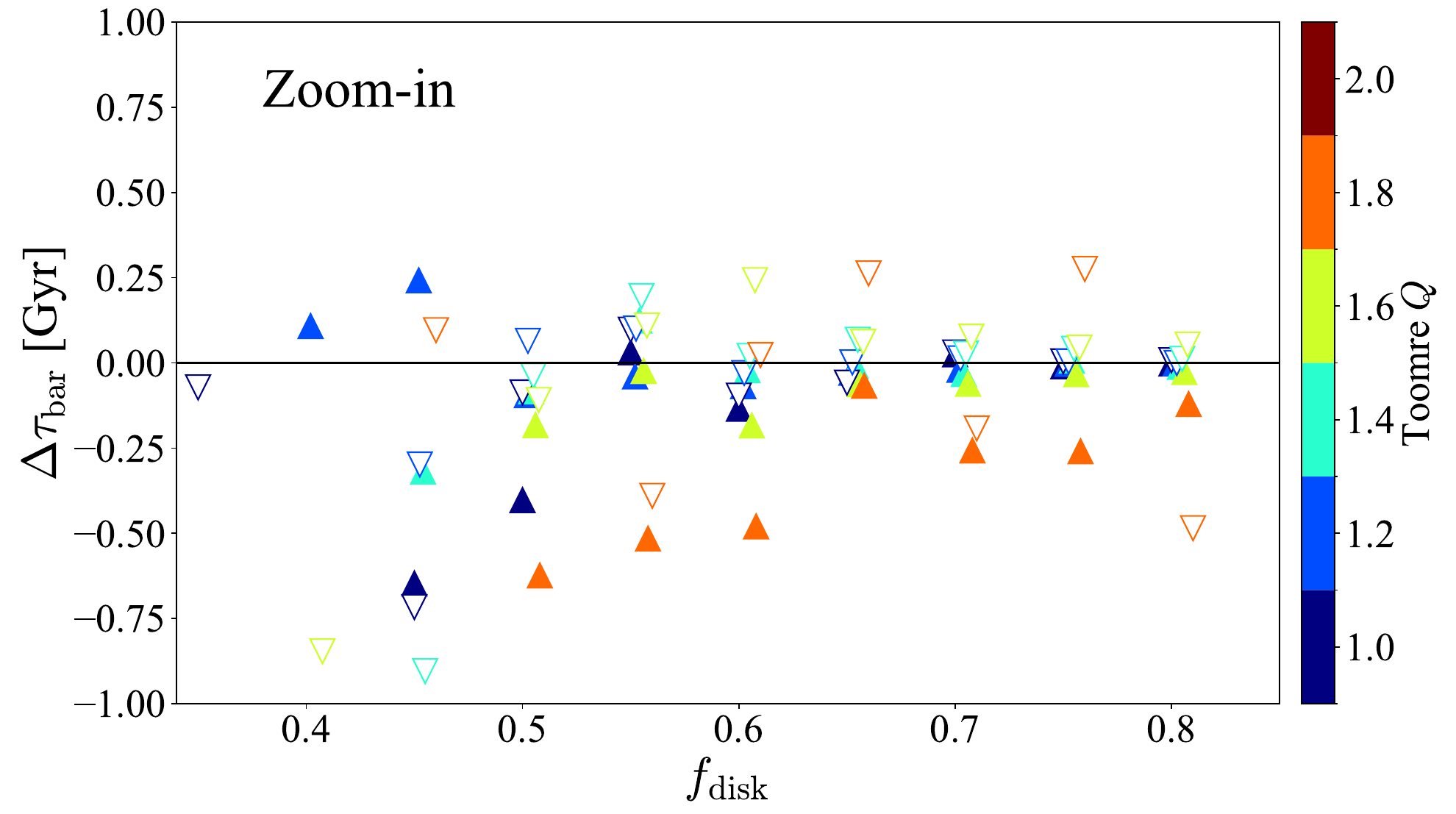}
\caption{Left panel: variation in bar formation timescale $\Delta\tbar$ as a function of $\fdisk$, where $\Delta\tbar$ is defined as the difference between the prograde (or retrograde) spin models and their zero-spin counterparts. Upward-filled triangles denote prograde models; downward unfilled triangles denote retrograde models. Only models with well-defined $\Delta\tbar$ values are included. Point colors indicate the Toomre $Q$ of each model. Small horizontal offsets are applied to reduce overlap among points with the same $\fdisk$. Right panel: zoom-in view of the left panel, limited to $|\Delta\tbar| \leq 1\ \mathrm{Gyr}$.}
\label{fig: tau bar variation}
\end{figure*}

\section{Results and Discussions}\label{sec: results}
\subsection{Impact of halo spin on $\tbar$}\label{sec: impact}

Figure~\ref{fig: spin effect} presents the time evolution of bar strength $A_2(t)$ for all models. Black curves correspond to zero-spin halos ($\lambda=0$), while green and red curves represent prograde and retrograde halo spin cases, respectively. Consistent with previous empirical trends reported in \cite{che_she_2025}, bar formation occurs earlier in models with higher $\fdisk$ and/or lower Toomre $Q$.

To more clearly assess the impact of halo spin on the bar formation process, we quantify the bar formation timescale for each model. Following \citet{bland_etal_2023} and \cite{che_she_2025}, we fit the initial exponential growth phase of $A_2(t)$ using Equation~\ref{eq: definition of tau bar}. The fitting is performed with \texttt{scipy.optimize.curve\_fit}, employing the default Levenberg–Marquardt algorithm to solve the nonlinear least-squares problem. The resulting fits are shown as dashed curves in Figure~\ref{fig: spin effect}. We apply the fitting only to models that undergo \emph{effective} bar formation, which we define empirically as $\operatorname{max}\{A_2(t)\} \geq 0.15$.

Figure~\ref{fig: Fujii diagram} shows the distribution of $\tbar$ as a function of $\fdisk$ for models exhibiting effective bar formation. The color of each data point indicates the corresponding Toomre $Q$ value. For comparison, we also plot the empirical relations from \citet{bland_etal_2023} and \cite{che_she_2025}. Overall, for both zero and prograde/retrograde spin halos, $\tbar$ follows a general trend of decreasing with increasing $\fdisk$, qualitatively consistent with previous findings. For the dependence on Toomre $Q$, although the trend deviates slightly from the global fit across the wider parameter space $0.2~\mathrm{kpc}\lesssim h_z\lesssim1.0~\mathrm{kpc}$, $\tbar$ still shows a secondary increase with increasing $Q$, consistent with the results of \citet{che_she_2025}.

In addition to the dependence on $\fdisk$ and Toomre $Q$, Figures~\ref{fig: spin effect} and \ref{fig: Fujii diagram} demonstrate that $\tbar$ also varies with halo spin. In models with intermediate parameters—specifically, $0.45 \lesssim \fdisk \lesssim 0.60$ and $Q \lesssim 1.8$—prograde spin is more likely to accelerate bar formation, while retrograde spin tends to slow it. This trend is consistent with previous findings by \citet{kat_she_2022}. However, the behavior differs for models with more extreme parameters: (1) At high $\fdisk$ and low $Q$ ($\fdisk \gtrsim 0.65$ and $Q \lesssim 1.4$), the differences in $\tbar$ among models with varying halo spin are negligible. (2) At low $\fdisk$, the effect of halo spin becomes more stochastic—whether prograde or retrograde spin promotes or suppresses bar formation varies from case to case, though the overall variation in $\tbar$ is more pronounced.

To directly visualize the impact of halo spin across different parameter regimes, Figure~\ref{fig: tau bar variation} shows the variation of $\tbar$ in spin-up models relative to their zero-spin counterparts. In the low-$\fdisk$ regime (left portion of the left panel), the variation in $\tbar$ is larger and more stochastic. In contrast, for models with higher $\fdisk$ (right panel), prograde spin generally accelerates bar formation, while retrograde spin slows it. The variation in $\tbar$ is more pronounced for intermediate $\fdisk$ than for the highest $\fdisk$ cases, where halo spin has only a minor effect.

The Appendix shows that these trends are independent of the particle number used for realizing the disk and halo. Together, they reflect the complex coupling between halo spin and bar formation.

\begin{figure*}
\centering
\includegraphics[width=.95\textwidth, height=.235\textheight]{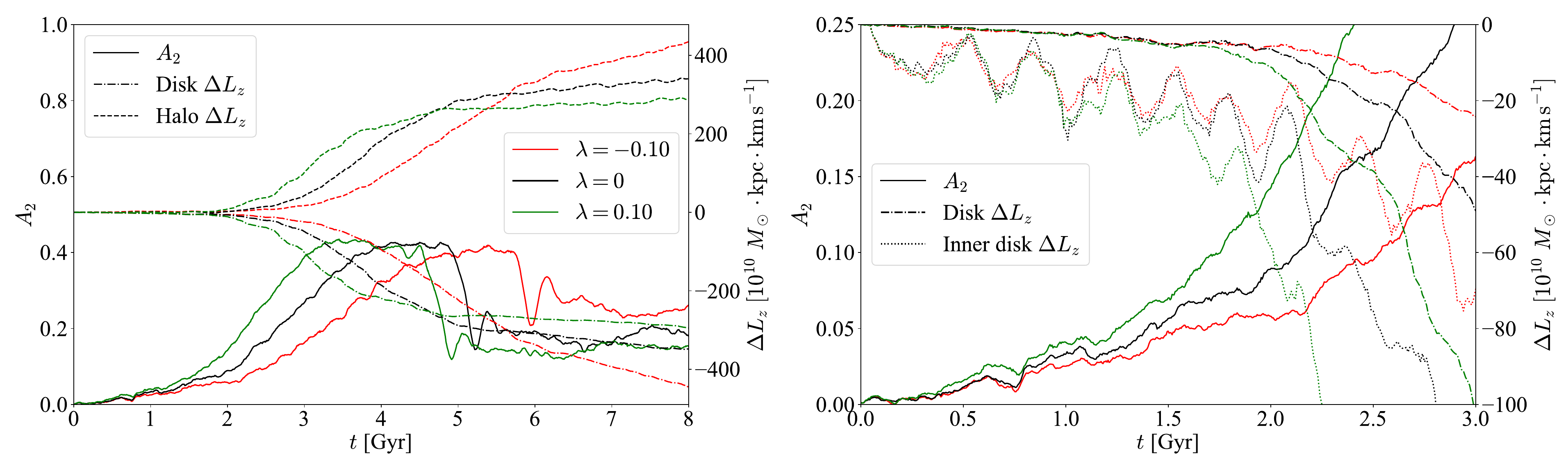}
\caption{Left panel: time evolution of $A_2$ (solid lines) for model $\multiSpinModel$ and its prograde and retrograde halo spin counterparts, with colors indicating halo spin. The corresponding variations of the vertical angular momentum $\Delta L_z$ of the disk (dotted-dashed lines) and DM halo (dashed lines) are also shown. Right panel: zoom-in on the early evolution; for better visualization, we restrict the range of $A_2$ to $[0,\ 0.25]$ and $\Delta{L}_z$ to $[-100,\ 0]$. Dotted lines show $\Delta{L}_z$ for the inner disk ($R<10\ \mathrm{kpc}$, $|z|<2.5\ \mathrm{kpc}$).}
\label{fig: A2 versus Lz}
\end{figure*}

\subsection{Bar formation and disk-halo $L_z$ transformation}

The previous section highlights the complex behavior of the bar formation timescale under varying halo spin. Angular momentum transfer between the stellar disk and DM halo is known to play a key role in bar evolution \citep{athana_2002, athana_2003_a, athana_2003_b, long_etal_2014}. To better understand how halo spin influences bar formation, we examine the angular momentum exchange in model $\multiSpinModel$, which exhibits a particularly strong spin-dependent response.

The left panel of Figure~\ref{fig: A2 versus Lz} shows the time evolution of bar strength $A_2$ alongside the variations of vertical angular momentum of the disk and halo. Around $t \sim 2\ \mathrm{Gyr}$, coincident with the rapid growth of $A_2$, a clear transfer of angular momentum from the disk to the halo is observed. The right panel focuses on the inner disk region—where the bar forms—and shows a stronger correlation between the onset of bar formation and the onset of $L_z$ exchange. As the halo spin varies, both the timing and shape of the $A_2$ and $\Delta{L}_z$ curves differ. Notably, although retrograde halos ultimately absorb more angular momentum \citep[consistent with previous studies; e.g.,][]{collie_etal_2018, collie_etal_2019, jan_kim_2024}, their initial $L_z$ absorption rate is slower. Since halo spin directly relates to the halo's initial angular momentum, these differences in $L_z$ evolution are naturally attributed to the spin. We therefore conclude that halo spin likely modulates the bar formation timescale by influencing the angular momentum transfer from the disk to the halo.

The angular momentum exchange framework offers a physical explanation for how halo spin can promote or slow bar formation. Halo absorbs angular momentum emitted by the inner disk mainly through resonances with the bar \citep{athana_2002, athana_2003_a}, which is more efficient if there are more halo particles participating in the resonances \citep{athana_2003_b}. In models with intermediate parameters—such as $\multiSpinModel$—the net rotation of a prograde spun-up halo is faster, so more particles are rotating at a speed close to the bar rotation, making them easier to be trapped into the resonances. As a result, the halo is more effective at absorbing the angular momentum shed from the inner disk during bar formation, thereby accelerating the onset of the bar. Conversely, in retrograde spin models, the decelerated halo rotation reduces the efficiency of angular momentum exchange, slowing bar formation. At higher $\fdisk$, bar formation is already rapid due to strong swing amplification \citep{che_she_2025}, leaving little room for halo spin to exert additional influence. As $\fdisk$ decreases or Toomre $Q$ increases, bar formation is slowed, allowing more room for the halo spin to affect the outcome. This explains why prograde halos more strongly promote bar formation moving from left to right across Figure~\ref{fig: spin effect}. In the lowest $\fdisk$ models, bar formation is significantly slowed, making the disk highly susceptible to the influence of halo spin, as seen in the large scatter of $\Delta\tbar$ in the left panel of Figure~\ref{fig: tau bar variation}. However, this increased sensitivity also makes bar formation more vulnerable to other perturbations and stochastic fluctuations, from both numerical noise in the simulations \citep{dubins_etal_2009} and nonlinear processes inherent to bar formation \citep{widrow_etal_2008}. As a result, these models do not show a systematic trend of prograde promotion or retrograde suppression, in contrast to the clearer trend observed at higher $\fdisk$.

\section{Summary} \label{sec: summary}

In this paper, consistent with previous studies, we confirm that lighter and/or hotter disks experience slowed bar formation, independent of halo spin.

We find that in the intermediate regime ($0.45 \lesssim \fdisk \lesssim 0.60$ and $1.0 \lesssim Q \lesssim 1.8$), prograde (retrograde) halo spin generally promotes (slows) bar formation, possibly by modulating the angular momentum exchange between the inner disk and the dark matter halo. For models with rapid bar formation ($\fdisk \gtrsim 0.65$ and $Q \lesssim 1.4$), the effect of halo spin is nearly negligible. In contrast, models with very long bar formation timescales are more susceptible to halo spin, resulting in large but stochastic variations in $\tbar$ when the halo is spun up.

These results may help explain the early emergence of bars in high-redshift galaxies, as observed by JWST \citep{guo_etal_2023, lecont_etal_2024}, potentially supported by prograde halo spin even in the existence or absence of significant gas content \citep{bland_etal_2023, bland_etal_2024, bland_etal_2025}.

\section*{acknowledgments}
We thank the anonymous referee for the careful review and helpful suggestions. BHC gratefully acknowledges the financial support from the China Scholarship Council and the support of the T.D. Lee scholarship. SKK acknowledges the support of the INSPIRE faculty grant (DST/INSPIRE/04/2023/000401) from the Department of Science \& Technology, India. The research presented here is partially supported by the National Natural Science Foundation of China under grant Nos. 12025302,  11773052, 11761131016; by China Manned Space Program with grant no. CMS-CSST-2025-A11; by the “111” Project of the Ministry of Education of China under grant No. B20019; by Shandong Provincial Key Research and Development Program (No. 2022CXGC020106); and by Pilot Project for Integrated Innovation of Science, Education and Industry of Qilu University of Technology (Shandong Academy of Sciences) (No. 2022JBZ01-01). This work made use of the Gravity Supercomputer at the Department of Astronomy, Shanghai Jiao Tong University, the facilities of the Center for High Performance Computing at Shanghai Astronomical Observatory, the facilities in the National Supercomputing Center in Jinan, and the Param Sanganak Supercomputing facilities at IIT Kanpur.

\appendix
\section{Independency test on the particle number} \label{app:indep}

\begin{figure*}
\centering
\includegraphics[width=.92\textwidth]{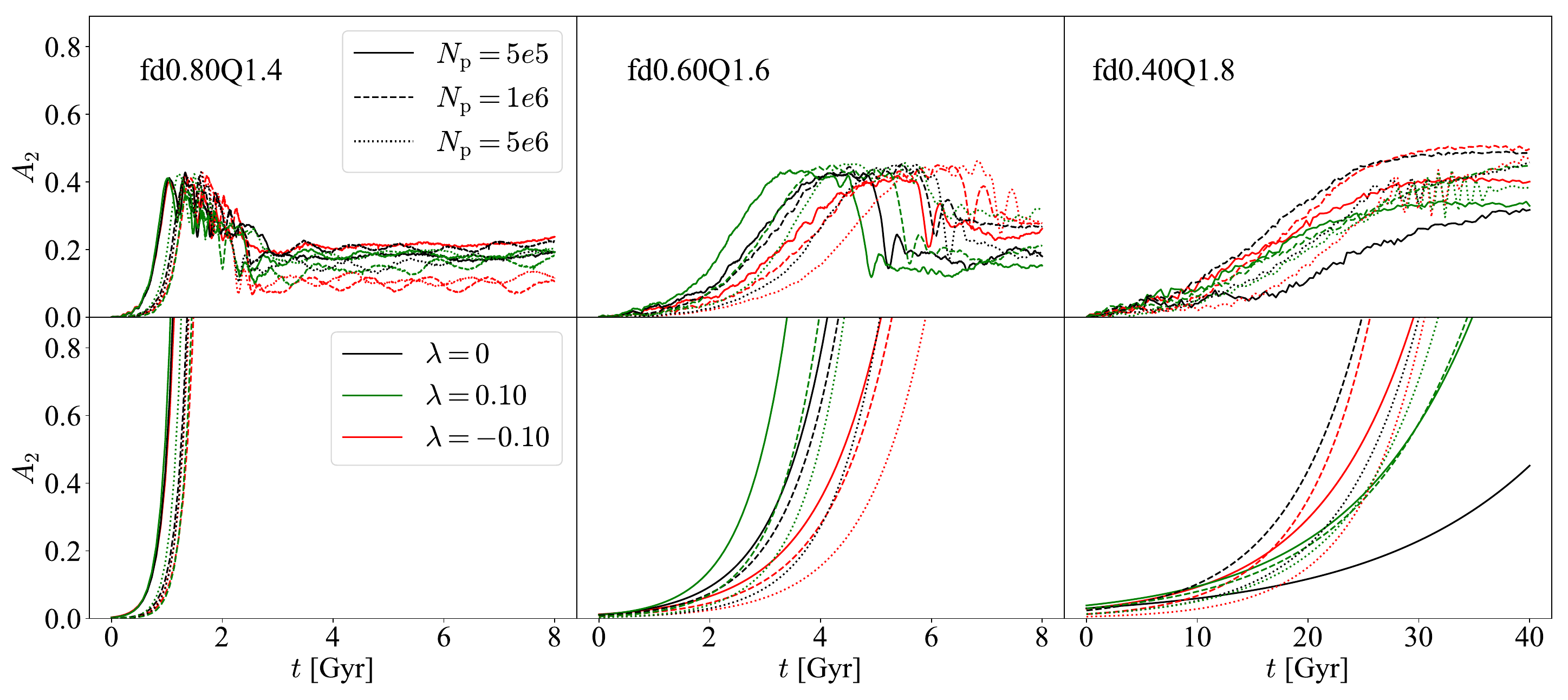}
\caption{First row: Time evolution of $A_2(t)$ for three models with $\lambda = 0$ (black), $\lambda = 0.10$ (green), and $\lambda = -0.10$ (red). Solid lines show the baseline models from Figure~\ref{fig: spin effect}; dashed and dotted lines correspond to models with two and ten times the number of particles ($N_\mathrm{p}$) in both the dark matter halo and stellar disk. Second row: exponential fits to the initial growth phase of $A_2(t)$ for each case.}
\label{fig: independent test}
\end{figure*}

\cite{che_she_2025} demonstrate that the empirical relation for bar formation timescales remains robust when the particle numbers are doubled. To further assess whether the number of particles affects the relationship between $\tbar$ and halo spin $\lambda$, we construct two higher-resolution models with two and ten times the number of particles ($N_\mathrm{p}$) used in the main sample. The results are shown in Figure~\ref{fig: independent test}. These three models represent the different regimes of halo spin influence discussed in Section~\ref{sec: results}: 

\begin{enumerate}[itemsep=0pt, topsep=3pt, parsep=0pt]
    \item $\mathrm{fd0.80Q1.4}$ — fast bar formation, where halo spin has nearly negligible effect;
    \item $\mathrm{fd0.60Q1.6}$ — intermediate bar formation, where prograde/retrograde spin tends to promote/slow bar onset;
    \item $\mathrm{fd0.40Q1.8}$ — slow bar formation, highly sensitive to halo spin and other perturbations, leading to significant but random (or nonmonotonic) variations in the bar formation timescale.
\end{enumerate}

As shown in Figure~\ref{fig: independent test}, the trends identified in the main text remain valid at higher particle resolution. This confirms that our results are robust against changes in $N_\mathrm{p}$.

\bibliography{references}{}

\begin{thebibliography}{}
\expandafter\ifx\csname natexlab\endcsname\relax\def\natexlab#1{#1}\fi
\providecommand{\url}[1]{\href{#1}{#1}}
\providecommand{\dodoi}[1]{doi:~\href{http://doi.org/#1}{\nolinkurl{#1}}}
\providecommand{\doeprint}[1]{\href{http://ascl.net/#1}{\nolinkurl{http://ascl.net/#1}}}
\providecommand{\doarXiv}[1]{\href{https://arxiv.org/abs/#1}{\nolinkurl{https://arxiv.org/abs/#1}}}

\bibitem[{S. {Ansar} \& M. {Das}(2024){Ansar} \& {Das}}]{Ansar.Das.2024}
{Ansar}, S., \& {Das}, M. 2024, \bibinfo{title}{{The Stellar
  Bar{\textendash}Dark Matter Halo Connection in the TNG50 Simulations},} \apj,
  975, 243, \dodoi{10.3847/1538-4357/ad7a6b}

\bibitem[{S. {Ansar} {et~al.}(2023){Ansar}, {Kataria}, \&
  {Das}}]{Ansar.et.al.2023}
{Ansar}, S., {Kataria}, S.~K., \& {Das}, M. 2023, \bibinfo{title}{{Modelling
  dark matter halo spin using observations and simulations: application to UGC
  5288},} \mnras, 522, 2967, \dodoi{10.1093/mnras/stad1060}

\bibitem[{E. {Athanassoula}(1992){Athanassoula}}]{athana_1992}
{Athanassoula}, E. 1992, \bibinfo{title}{{The existence and shapes of dust
  lanes in galactic bars.},} \mnras, 259, 345, \dodoi{10.1093/mnras/259.2.345}

\bibitem[{E. {Athanassoula}(2002){Athanassoula}}]{athana_2002}
{Athanassoula}, E. 2002, \bibinfo{title}{{Bar-Halo Interaction and Bar
  Growth},} \apjl, 569, L83, \dodoi{10.1086/340784}

\bibitem[{E. {Athanassoula}(2003{\natexlab{a}}){Athanassoula}}]{athana_2003_a}
{Athanassoula}, E. 2003{\natexlab{a}}, in Galaxies and Chaos, ed.
  G.~{Contopoulos} \& N.~{Voglis}, Vol. 626, 313--326,
  \dodoi{10.1007/978-3-540-45040-5_26}

\bibitem[{E. {Athanassoula}(2003{\natexlab{b}}){Athanassoula}}]{athana_2003_b}
{Athanassoula}, E. 2003{\natexlab{b}}, \bibinfo{title}{{What determines the
  strength and the slowdown rate of bars?},} \mnras, 341, 1179,
  \dodoi{10.1046/j.1365-8711.2003.06473.x}

\bibitem[{E. {Athanassoula}(2005){Athanassoula}}]{athana_2005}
{Athanassoula}, E. 2005, \bibinfo{title}{{On the nature of bulges in general
  and of box/peanut bulges in particular: input from N-body simulations},}
  \mnras, 358, 1477, \dodoi{10.1111/j.1365-2966.2005.08872.x}

\bibitem[{E. {Athanassoula} \& J.~A. {Sellwood}(1986){Athanassoula} \&
  {Sellwood}}]{ath_sel_1986}
{Athanassoula}, E., \& {Sellwood}, J.~A. 1986, \bibinfo{title}{{Bi-symmetric
  instabilities of the Kuz'min/Toomre disc.},} \mnras, 221, 213,
  \dodoi{10.1093/mnras/221.2.213}

\bibitem[{F.~D. {Barazza} {et~al.}(2008){Barazza}, {Jogee}, \&
  {Marinova}}]{barazz_etal_2008}
{Barazza}, F.~D., {Jogee}, S., \& {Marinova}, I. 2008, \bibinfo{title}{{Bars in
  Disk-dominated and Bulge-dominated Galaxies at z
  \raisebox{-0.5ex}\textasciitilde 0: New Insights from
  \raisebox{-0.5ex}\textasciitilde3600 SDSS Galaxies},} \apj, 675, 1194,
  \dodoi{10.1086/526510}

\bibitem[{J. {Barnes} \& G. {Efstathiou}(1987){Barnes} \&
  {Efstathiou}}]{bar_efs_1987}
{Barnes}, J., \& {Efstathiou}, G. 1987, \bibinfo{title}{{Angular Momentum from
  Tidal Torques},} \apj, 319, 575, \dodoi{10.1086/165480}

\bibitem[{L. {Beraldo e Silva} {et~al.}(2023){Beraldo e Silva}, {Debattista},
  {Anderson}, {Valluri}, {Erwin}, {Daniel}, \& {Deg}}]{berald_etal_2023}
{Beraldo e Silva}, L., {Debattista}, V.~P., {Anderson}, S.~R., {et~al.} 2023,
  \bibinfo{title}{{Orbital Support and Evolution of Flat Profiles of Bars
  (Shoulders)},} \apj, 955, 38, \dodoi{10.3847/1538-4357/ace976}

\bibitem[{P. {Bett} {et~al.}(2007){Bett}, {Eke}, {Frenk}, {Jenkins}, {Helly},
  \& {Navarro}}]{bett_etal_2007}
{Bett}, P., {Eke}, V., {Frenk}, C.~S., {et~al.} 2007, \bibinfo{title}{{The spin
  and shape of dark matter haloes in the Millennium simulation of a
  {\ensuremath{\Lambda}} cold dark matter universe},} \mnras, 376, 215,
  \dodoi{10.1111/j.1365-2966.2007.11432.x}

\bibitem[{J. {Binney}(2020){Binney}}]{binney_2020}
{Binney}, J. 2020, \bibinfo{title}{{The shearing sheet and swing amplification
  revisited},} \mnras, 496, 767, \dodoi{10.1093/mnras/staa1485}

\bibitem[{J. {Bland-Hawthorn} {et~al.}(2024){Bland-Hawthorn}, {Tepper-Garcia},
  {Agertz}, \& {Federrath}}]{bland_etal_2024}
{Bland-Hawthorn}, J., {Tepper-Garcia}, T., {Agertz}, O., \& {Federrath}, C.
  2024, \bibinfo{title}{{Turbulent Gas-rich Disks at High Redshift: Bars and
  Bulges in a Radial Shear Flow},} \apj, 968, 86,
  \dodoi{10.3847/1538-4357/ad4118}

\bibitem[{J. {Bland-Hawthorn} {et~al.}(2023){Bland-Hawthorn}, {Tepper-Garcia},
  {Agertz}, \& {Freeman}}]{bland_etal_2023}
{Bland-Hawthorn}, J., {Tepper-Garcia}, T., {Agertz}, O., \& {Freeman}, K. 2023,
  \bibinfo{title}{{The Rapid Onset of Stellar Bars in the Baryon-dominated
  Centers of Disk Galaxies},} \apj, 947, 80, \dodoi{10.3847/1538-4357/acc469}

\bibitem[{J. {Bland-Hawthorn} {et~al.}(2025){Bland-Hawthorn}, {Tepper-Garcia},
  {Agertz}, {Federrath}, {Haywood}, {di Matteo}, {Bedding}, {Tsukui},
  {Wisnioski}, {Ness}, \& {Freeman}}]{bland_etal_2025}
{Bland-Hawthorn}, J., {Tepper-Garcia}, T., {Agertz}, O., {et~al.} 2025,
  \bibinfo{title}{{Turbulent gas-rich discs at high redshift: origin of thick
  stellar discs through 3D 'baryon sloshing'},} arXiv e-prints,
  arXiv:2502.01895, \dodoi{10.48550/arXiv.2502.01895}

\bibitem[{J.~S. {Bullock} {et~al.}(2001){Bullock}, {Dekel}, {Kolatt},
  {Kravtsov}, {Klypin}, {Porciani}, \& {Primack}}]{bulloc_etal_2001}
{Bullock}, J.~S., {Dekel}, A., {Kolatt}, T.~S., {et~al.} 2001,
  \bibinfo{title}{{A Universal Angular Momentum Profile for Galactic Halos},}
  \apj, 555, 240, \dodoi{10.1086/321477}

\bibitem[{B.-H. {Chen} \& Z.-Y. {Li}(2022){Chen} \& {Li}}]{che_li_2022}
{Chen}, B.-H., \& {Li}, Z.-Y. 2022, \bibinfo{title}{{Metallicity Properties of
  the Galactic Bulge Stars Near and Far: Expectations from the Auriga
  Simulation},} \apj, 934, 28, \dodoi{10.3847/1538-4357/ac795c}

\bibitem[{B.-H. {Chen} \& J. {Shen}(2025){Chen} \& {Shen}}]{che_she_2025}
{Chen}, B.-H., \& {Shen}, J. 2025, \bibinfo{title}{{The Dependency of Bar
  Formation Timescale on Disk Mass Fraction, Toomre Q, and Scale Height},}
  \apj, 990, 140, \dodoi{10.3847/1538-4357/adf966}

\bibitem[{B.-H. {Chen} {et~al.}(2024){Chen}, {Shen}, \& {Liu}}]{chen_etal_2024}
{Chen}, B.-H., {Shen}, J., \& {Liu}, Z. 2024, \bibinfo{title}{{Dynamical Origin
  of the Vertical Metallicity Gradient of the Milky Way Bulge},} \apj, 976,
  232, \dodoi{10.3847/1538-4357/ad8640}

\bibitem[{R. {Chiba} \& S.~K. {Kataria}(2024){Chiba} \&
  {Kataria}}]{chi_kat_2024}
{Chiba}, R., \& {Kataria}, S.~K. 2024, \bibinfo{title}{{Origin of reduced
  dynamical friction by dark matter haloes with net prograde rotation},}
  \mnras, 528, 4115, \dodoi{10.1093/mnras/stae288}

\bibitem[{A. {Collier} {et~al.}(2018){Collier}, {Shlosman}, \&
  {Heller}}]{collie_etal_2018}
{Collier}, A., {Shlosman}, I., \& {Heller}, C. 2018, \bibinfo{title}{{What
  makes the family of barred disc galaxies so rich: damping stellar bars in
  spinning haloes},} \mnras, 476, 1331, \dodoi{10.1093/mnras/sty270}

\bibitem[{A. {Collier} {et~al.}(2019){Collier}, {Shlosman}, \&
  {Heller}}]{collie_etal_2019}
{Collier}, A., {Shlosman}, I., \& {Heller}, C. 2019, \bibinfo{title}{{Stellar
  bars in counter-rotating dark matter haloes: the role of halo orbit
  reversals},} \mnras, 489, 3102, \dodoi{10.1093/mnras/stz2327}

\bibitem[{F. {Combes} {et~al.}(1990){Combes}, {Debbasch}, {Friedli}, \&
  {Pfenniger}}]{combes_etal_1990}
{Combes}, F., {Debbasch}, F., {Friedli}, D., \& {Pfenniger}, D. 1990,
  \bibinfo{title}{{Box and peanut shapes generated by stellar bars.},} \aap,
  233, 82

\bibitem[{F. {Combes} \& R.~H. {Sanders}(1981){Combes} \&
  {Sanders}}]{com_san_1981}
{Combes}, F., \& {Sanders}, R.~H. 1981, \bibinfo{title}{{Formation and
  properties of persisting stellar bars.},} \aap, 96, 164

\bibitem[{V.~P. {Debattista} {et~al.}(2020){Debattista}, {Liddicott},
  {Khachaturyants}, \& {Beraldo e Silva}}]{debatt_etal_2020}
{Debattista}, V.~P., {Liddicott}, D.~J., {Khachaturyants}, T., \& {Beraldo e
  Silva}, L. 2020, \bibinfo{title}{{Box/peanut-shaped bulges in action space},}
  \mnras, 498, 3334, \dodoi{10.1093/mnras/staa2568}

\bibitem[{V.~P. {Debattista} {et~al.}(2017){Debattista}, {Ness}, {Gonzalez},
  {Freeman}, {Zoccali}, \& {Minniti}}]{debatt_etal_2017}
{Debattista}, V.~P., {Ness}, M., {Gonzalez}, O.~A., {et~al.} 2017,
  \bibinfo{title}{{Separation of stellar populations by an evolving bar:
  implications for the bulge of the Milky Way},} \mnras, 469, 1587,
  \dodoi{10.1093/mnras/stx947}

\bibitem[{A.~G. {Doroshkevich}(1970){Doroshkevich}}]{dorosh_1970}
{Doroshkevich}, A.~G. 1970, \bibinfo{title}{{The space structure of
  perturbations and the origin of rotation of galaxies in the theory of
  fluctuation.},} Astrofizika, 6, 581

\bibitem[{J. {Dubinski} {et~al.}(2009){Dubinski}, {Berentzen}, \&
  {Shlosman}}]{dubins_etal_2009}
{Dubinski}, J., {Berentzen}, I., \& {Shlosman}, I. 2009,
  \bibinfo{title}{{Anatomy of the Bar Instability in Cuspy Dark Matter Halos},}
  \apj, 697, 293, \dodoi{10.1088/0004-637X/697/1/293}

\bibitem[{G. {Efstathiou} {et~al.}(1982){Efstathiou}, {Lake}, \&
  {Negroponte}}]{efstat_etal_1982}
{Efstathiou}, G., {Lake}, G., \& {Negroponte}, J. 1982, \bibinfo{title}{{The
  stability and masses of disc galaxies.},} \mnras, 199, 1069,
  \dodoi{10.1093/mnras/199.4.1069}

\bibitem[{P. {Erwin}(2018){Erwin}}]{erwin_2018}
{Erwin}, P. 2018, \bibinfo{title}{{The dependence of bar frequency on galaxy
  mass, colour, and gas content - and angular resolution - in the local
  universe},} \mnras, 474, 5372, \dodoi{10.1093/mnras/stx3117}

\bibitem[{P.~B. {Eskridge} {et~al.}(2000){Eskridge}, {Frogel}, {Pogge},
  {Quillen}, {Davies}, {DePoy}, {Houdashelt}, {Kuchinski}, {Ram{\'\i}rez},
  {Sellgren}, {Terndrup}, \& {Tiede}}]{eskrid_etal_2000}
{Eskridge}, P.~B., {Frogel}, J.~A., {Pogge}, R.~W., {et~al.} 2000,
  \bibinfo{title}{{The Frequency of Barred Spiral Galaxies in the
  Near-Infrared},} \aj, 119, 536, \dodoi{10.1086/301203}

\bibitem[{F. {Fragkoudi} {et~al.}(2025){Fragkoudi}, {Grand}, {Pakmor},
  {G{\'o}mez}, {Marinacci}, \& {Springel}}]{fragko_etal_2025}
{Fragkoudi}, F., {Grand}, R. J.~J., {Pakmor}, R., {et~al.} 2025,
  \bibinfo{title}{{Bar formation and evolution in the cosmological context:
  inputs from the Auriga simulations},} \mnras, 538, 1587,
  \dodoi{10.1093/mnras/staf389}

\bibitem[{F. {Fragkoudi} {et~al.}(2019){Fragkoudi}, {Katz}, {Trick}, {White},
  {Di Matteo}, {Sormani}, {Khoperskov}, {Haywood}, {Hall{\'e}}, \&
  {G{\'o}mez}}]{fragko_etal_2019}
{Fragkoudi}, F., {Katz}, D., {Trick}, W., {et~al.} 2019, \bibinfo{title}{{On
  the ridges, undulations, and streams in Gaia DR2: linking the topography of
  phase space to the orbital structure of an N-body bar},} \mnras, 488, 3324,
  \dodoi{10.1093/mnras/stz1875}

\bibitem[{M.~S. {Fujii} {et~al.}(2018){Fujii}, {B{\'e}dorf}, {Baba}, \&
  {Portegies Zwart}}]{fujii_etal_2018}
{Fujii}, M.~S., {B{\'e}dorf}, J., {Baba}, J., \& {Portegies Zwart}, S. 2018,
  \bibinfo{title}{{The dynamics of stellar discs in live dark-matter haloes},}
  \mnras, 477, 1451, \dodoi{10.1093/mnras/sty711}

\bibitem[{D.~A. {Gadotti} {et~al.}(2020){Gadotti}, {Bittner},
  {Falc{\'o}n-Barroso}, {M{\'e}ndez-Abreu}, {Kim}, {Fragkoudi}, {de
  Lorenzo-C{\'a}ceres}, {Leaman}, {Neumann}, {Querejeta},
  {S{\'a}nchez-Bl{\'a}zquez}, {Martig}, {Mart{\'\i}n-Navarro}, {P{\'e}rez},
  {Seidel}, \& {van de Ven}}]{gadott_etal_2020}
{Gadotti}, D.~A., {Bittner}, A., {Falc{\'o}n-Barroso}, J., {et~al.} 2020,
  \bibinfo{title}{{Kinematic signatures of nuclear discs and bar-driven secular
  evolution in nearby galaxies of the MUSE TIMER project},} \aap, 643, A14,
  \dodoi{10.1051/0004-6361/202038448}

\bibitem[{R.~J.~J. {Grand} {et~al.}(2024){Grand}, {Fragkoudi}, {G{\'o}mez},
  {Jenkins}, {Marinacci}, {Pakmor}, \& {Springel}}]{grand_etal_2024}
{Grand}, R. J.~J., {Fragkoudi}, F., {G{\'o}mez}, F.~A., {et~al.} 2024,
  \bibinfo{title}{{Overview and public data release of the augmented Auriga
  Project: cosmological simulations of dwarf and Milky Way-mass galaxies},}
  \mnras, 532, 1814, \dodoi{10.1093/mnras/stae1598}

\bibitem[{R.~J.~J. {Grand} {et~al.}(2017){Grand}, {G{\'o}mez}, {Marinacci},
  {Pakmor}, {Springel}, {Campbell}, {Frenk}, {Jenkins}, \&
  {White}}]{grand_etal_2017}
{Grand}, R. J.~J., {G{\'o}mez}, F.~A., {Marinacci}, F., {et~al.} 2017,
  \bibinfo{title}{{The Auriga Project: the properties and formation mechanisms
  of disc galaxies across cosmic time},} \mnras, 467, 179,
  \dodoi{10.1093/mnras/stx071}

\bibitem[{Y. {Guo} {et~al.}(2023){Guo}, {Jogee}, {Finkelstein}, {Chen}, {Wise},
  {Bagley}, {Barro}, {Wuyts}, {Kocevski}, {Kartaltepe}, {McGrath}, {Ferguson},
  {Mobasher}, {Giavalisco}, {Lucas}, {Zavala}, {Lotz}, {Grogin},
  {Huertas-Company}, {Vega-Ferrero}, {Hathi}, {Arrabal Haro}, {Dickinson},
  {Koekemoer}, {Papovich}, {Pirzkal}, {Yung}, {Backhaus}, {Bell},
  {Calabr{\`o}}, {Cleri}, {Coogan}, {Cooper}, {Costantin}, {Croton}, {Davis},
  {Dekel}, {Franco}, {Gardner}, {Holwerda}, {Hutchison}, {Pandya},
  {P{\'e}rez-Gonz{\'a}lez}, {Ravindranath}, {Rose}, {Trump}, {de la Vega}, \&
  {Wang}}]{guo_etal_2023}
{Guo}, Y., {Jogee}, S., {Finkelstein}, S.~L., {et~al.} 2023,
  \bibinfo{title}{{First Look at z > 1 Bars in the Rest-frame Near-infrared
  with JWST Early CEERS Imaging},} \apjl, 945, L10,
  \dodoi{10.3847/2041-8213/acacfb}

\bibitem[{H. {Hetznecker} \& A. {Burkert}(2006){Hetznecker} \&
  {Burkert}}]{het_bur_2006}
{Hetznecker}, H., \& {Burkert}, A. 2006, \bibinfo{title}{{The evolution of the
  dark halo spin parameters {\ensuremath{\lambda}} and {\ensuremath{\lambda}}'
  in a {\ensuremath{\Lambda}}CDM universe: the role of minor and major
  mergers},} \mnras, 370, 1905, \dodoi{10.1111/j.1365-2966.2006.10616.x}

\bibitem[{F. {Hohl}(1971){Hohl}}]{hohl_1971}
{Hohl}, F. 1971, \bibinfo{title}{{Numerical Experiments with a Disk of Stars},}
  \apj, 168, 343, \dodoi{10.1086/151091}

\bibitem[{F. {Hoyle}(1949){Hoyle}}]{hoyle_1949}
{Hoyle}, F. 1949, \bibinfo{title}{{On the Cosmological Problem},} \mnras, 109,
  365, \dodoi{10.1093/mnras/109.3.365}

\bibitem[{T. {Ishiyama} {et~al.}(2013){Ishiyama}, {Rieder}, {Makino},
  {Portegies Zwart}, {Groen}, {Nitadori}, {de Laat}, {McMillan}, {Hiraki}, \&
  {Harfst}}]{ishiya_etal_2013}
{Ishiyama}, T., {Rieder}, S., {Makino}, J., {et~al.} 2013, \bibinfo{title}{{The
  Cosmogrid Simulation: Statistical Properties of Small Dark Matter Halos},}
  \apj, 767, 146, \dodoi{10.1088/0004-637X/767/2/146}

\bibitem[{D. {Jang} \& W.-T. {Kim}(2023){Jang} \& {Kim}}]{jan_kim_2023}
{Jang}, D., \& {Kim}, W.-T. 2023, \bibinfo{title}{{Effects of the Central Mass
  Concentration on Bar Formation in Disk Galaxies},} \apj, 942, 106,
  \dodoi{10.3847/1538-4357/aca7bc}

\bibitem[{D. {Jang} \& W.-T. {Kim}(2024){Jang} \& {Kim}}]{jan_kim_2024}
{Jang}, D., \& {Kim}, W.-T. 2024, \bibinfo{title}{{Effects of Halo Spin on the
  Formation and Evolution of Bars in Disk Galaxies},} \apj, 971, 67,
  \dodoi{10.3847/1538-4357/ad54b9}

\bibitem[{S.~K. {Kataria}(2024){Kataria}}]{Kataria.2024}
{Kataria}, S.~K. 2024, \bibinfo{title}{{How do the successive buckling events
  affect a galaxy bar and stellar disc? Potential observable signatures for
  spotting the buckling action - I},} \mnras, 534, 3565,
  \dodoi{10.1093/mnras/stae2311}

\bibitem[{S.~K. {Kataria} \& M. {Das}(2018){Kataria} \& {Das}}]{kat_das_2018}
{Kataria}, S.~K., \& {Das}, M. 2018, \bibinfo{title}{{A study of the effect of
  bulges on bar formation in disc galaxies},} \mnras, 475, 1653,
  \dodoi{10.1093/mnras/stx3279}

\bibitem[{S.~K. {Kataria} \& M. {Das}(2019){Kataria} \& {Das}}]{kat_das_2019}
{Kataria}, S.~K., \& {Das}, M. 2019, \bibinfo{title}{{The Effect of Bulge Mass
  on Bar Pattern Speed in Disk Galaxies},} \apj, 886, 43,
  \dodoi{10.3847/1538-4357/ab48f7}

\bibitem[{S.~K. {Kataria} {et~al.}(2020){Kataria}, {Das}, \&
  {Barway}}]{katari_etal_2020}
{Kataria}, S.~K., {Das}, M., \& {Barway}, S. 2020, \bibinfo{title}{{Testing a
  theoretical prediction for bar formation in galaxies with bulges},} \aap,
  640, A14, \dodoi{10.1051/0004-6361/202037527}

\bibitem[{S.~K. {Kataria} \& J. {Shen}(2022){Kataria} \& {Shen}}]{kat_she_2022}
{Kataria}, S.~K., \& {Shen}, J. 2022, \bibinfo{title}{{Effects of Inner Halo
  Angular Momentum on the Peanut/X Shapes of Bars},} \apj, 940, 175,
  \dodoi{10.3847/1538-4357/ac9df1}

\bibitem[{S.~K. {Kataria} \& J. {Shen}(2024){Kataria} \& {Shen}}]{kat_she_2024}
{Kataria}, S.~K., \& {Shen}, J. 2024, \bibinfo{title}{{Importance of Initial
  Condition on Bar Secular Evolution: Role of Halo Angular Momentum
  Distribution Discontinuity},} \apj, 970, 45, \dodoi{10.3847/1538-4357/ad5b58}

\bibitem[{S.~K. {Kataria} \& M. {Vivek}(2024){Kataria} \&
  {Vivek}}]{kat_viv_2024}
{Kataria}, S.~K., \& {Vivek}, M. 2024, \bibinfo{title}{{How does the presence
  of bar affects the fueling of supermassive black holes? An IllustrisTNG100
  perspective},} \mnras, 527, 3366, \dodoi{10.1093/mnras/stad3383}

\bibitem[{T. {Kim} {et~al.}(2024){Kim}, {Gadotti}, {Lee}, {L{\'o}pez-Cob{\'a}},
  {Kim}, {Kim}, \& {Park}}]{kim_etal_2024}
{Kim}, T., {Gadotti}, D.~A., {Lee}, Y.~H., {et~al.} 2024, \bibinfo{title}{{Do
  Strong Bars Exhibit Strong Noncircular Motions?},} \apj, 976, 220,
  \dodoi{10.3847/1538-4357/ad8573}

\bibitem[{A. {Knebe} \& C. {Power}(2008){Knebe} \& {Power}}]{kne_pow_2008}
{Knebe}, A., \& {Power}, C. 2008, \bibinfo{title}{{On the Correlation between
  Spin Parameter and Halo Mass},} \apj, 678, 621, \dodoi{10.1086/586702}

\bibitem[{J. {Kormendy} \& J. {Kennicutt}(2004){Kormendy} \&
  {Kennicutt}}]{kor_ken_2004}
{Kormendy}, J., \& {Kennicutt}, Robert~C., J. 2004, \bibinfo{title}{{Secular
  Evolution and the Formation of Pseudobulges in Disk Galaxies},} \araa, 42,
  603, \dodoi{10.1146/annurev.astro.42.053102.134024}

\bibitem[{A. {Kumar} {et~al.}(2022){Kumar}, {Das}, \&
  {Kataria}}]{Kumar.et.al.2022}
{Kumar}, A., {Das}, M., \& {Kataria}, S.~K. 2022, \bibinfo{title}{{The effect
  of dark matter halo shape on bar buckling and boxy/peanut bulges},} \mnras,
  509, 1262, \dodoi{10.1093/mnras/stab3019}

\bibitem[{Z.~A. {Le Conte} {et~al.}(2024){Le Conte}, {Gadotti}, {Ferreira},
  {Conselice}, {de S{\'a}-Freitas}, {Kim}, {Neumann}, {Fragkoudi},
  {Athanassoula}, \& {Adams}}]{lecont_etal_2024}
{Le Conte}, Z.~A., {Gadotti}, D.~A., {Ferreira}, L., {et~al.} 2024,
  \bibinfo{title}{{A JWST investigation into the bar fraction at redshifts 1
  {\ensuremath{\leq}} z {\ensuremath{\leq}} 3},} \mnras, 530, 1984,
  \dodoi{10.1093/mnras/stae921}

\bibitem[{Y.~H. {Lee} {et~al.}(2019){Lee}, {Ann}, \& {Park}}]{lee_etal_2019}
{Lee}, Y.~H., {Ann}, H.~B., \& {Park}, M.-G. 2019, \bibinfo{title}{{Bar
  Fraction in Early- and Late-type Spirals},} \apj, 872, 97,
  \dodoi{10.3847/1538-4357/ab0024}

\bibitem[{Z. {Li} {et~al.}(2016){Li}, {Gerhard}, {Shen}, {Portail}, \&
  {Wegg}}]{li_etal_2016}
{Li}, Z., {Gerhard}, O., {Shen}, J., {Portail}, M., \& {Wegg}, C. 2016,
  \bibinfo{title}{{Gas Dynamics in the Milky Way: A Low Pattern Speed Model},}
  \apj, 824, 13, \dodoi{10.3847/0004-637X/824/1/13}

\bibitem[{Z. {Li} {et~al.}(2017){Li}, {Sellwood}, \& {Shen}}]{li_etal_2017}
{Li}, Z., {Sellwood}, J.~A., \& {Shen}, J. 2017, \bibinfo{title}{{Rapid
  Formation of Black Holes in Galaxies: A Self-limiting Growth Mechanism},}
  \apj, 850, 67, \dodoi{10.3847/1538-4357/aa9377}

\bibitem[{Z. {Li} {et~al.}(2023){Li}, {Du}, {Debattista}, {Shen}, {Li}, {Liu},
  {Vogelsberger}, {Beane}, {Marinacci}, \& {Sales}}]{li_etal_2023}
{Li}, Z., {Du}, M., {Debattista}, V.~P., {et~al.} 2023, \bibinfo{title}{{How
  Nested Bars Enhance, Modulate, and Are Destroyed by Gas Inflows},} \apj, 958,
  77, \dodoi{10.3847/1538-4357/acffb3}

\bibitem[{Z.-Y. {Li} \& J. {Shen}(2012){Li} \& {Shen}}]{li_she_2012}
{Li}, Z.-Y., \& {Shen}, J. 2012, \bibinfo{title}{{The Vertical X-shaped
  Structure in the Milky Way: Evidence from a Simple Boxy Bulge Model},} \apjl,
  757, L7, \dodoi{10.1088/2041-8205/757/1/L7}

\bibitem[{Z.-Y. {Li} \& J. {Shen}(2015){Li} \& {Shen}}]{li_she_2015}
{Li}, Z.-Y., \& {Shen}, J. 2015, \bibinfo{title}{{Mapping the Three-dimensional
  ``X-shaped Structure'' in Models of the Galactic Bulge},} \apjl, 815, L20,
  \dodoi{10.1088/2041-8205/815/2/L20}

\bibitem[{S. {Long} {et~al.}(2014){Long}, {Shlosman}, \&
  {Heller}}]{long_etal_2014}
{Long}, S., {Shlosman}, I., \& {Heller}, C. 2014, \bibinfo{title}{{Secular
  Damping of Stellar Bars in Spinning Dark Matter Halos},} \apjl, 783, L18,
  \dodoi{10.1088/2041-8205/783/1/L18}

\bibitem[{D. {Lynden-Bell} \& A.~J. {Kalnajs}(1972){Lynden-Bell} \&
  {Kalnajs}}]{lyn_kal_1972}
{Lynden-Bell}, D., \& {Kalnajs}, A.~J. 1972, \bibinfo{title}{{On the generating
  mechanism of spiral structure},} \mnras, 157, 1,
  \dodoi{10.1093/mnras/157.1.1}

\bibitem[{A.~H. {Maller} {et~al.}(2002){Maller}, {Dekel}, \&
  {Somerville}}]{maller_etal_2002}
{Maller}, A.~H., {Dekel}, A., \& {Somerville}, R. 2002,
  \bibinfo{title}{{Modelling angular-momentum history in dark-matter haloes},}
  \mnras, 329, 423, \dodoi{10.1046/j.1365-8711.2002.04983.x}

\bibitem[{K. {Men{\'e}ndez-Delmestre} {et~al.}(2007){Men{\'e}ndez-Delmestre},
  {Sheth}, {Schinnerer}, {Jarrett}, \& {Scoville}}]{menend_etal_2007}
{Men{\'e}ndez-Delmestre}, K., {Sheth}, K., {Schinnerer}, E., {Jarrett}, T.~H.,
  \& {Scoville}, N.~Z. 2007, \bibinfo{title}{{A Near-Infrared Study of 2MASS
  Bars in Local Galaxies: An Anchor for High-Redshift Studies},} \apj, 657,
  790, \dodoi{10.1086/511025}

\bibitem[{I. {Minchev} \& B. {Famaey}(2010){Minchev} \&
  {Famaey}}]{min_fam_2010}
{Minchev}, I., \& {Famaey}, B. 2010, \bibinfo{title}{{A New Mechanism for
  Radial Migration in Galactic Disks: Spiral-Bar Resonance Overlap},} \apj,
  722, 112, \dodoi{10.1088/0004-637X/722/1/112}

\bibitem[{J.~F. {Navarro} {et~al.}(1996){Navarro}, {Frenk}, \&
  {White}}]{navarr_etal_1996}
{Navarro}, J.~F., {Frenk}, C.~S., \& {White}, S. D.~M. 1996,
  \bibinfo{title}{{The Structure of Cold Dark Matter Halos},} \apj, 462, 563,
  \dodoi{10.1086/177173}

\bibitem[{J. {Neumann} {et~al.}(2024){Neumann}, {Thomas}, {Maraston}, {Gleis},
  {Mao}, {Schinnerer}, \& {Stuber}}]{neuman_etal_2024}
{Neumann}, J., {Thomas}, D., {Maraston}, C., {et~al.} 2024,
  \bibinfo{title}{{Azimuthal variations of stellar populations in barred
  galaxies},} \mnras, 534, 2438, \dodoi{10.1093/mnras/stae2252}

\bibitem[{P.~J.~E. {Peebles}(1969){Peebles}}]{peeble_1969}
{Peebles}, P.~J.~E. 1969, \bibinfo{title}{{Origin of the Angular Momentum of
  Galaxies},} \apj, 155, 393, \dodoi{10.1086/149876}

\bibitem[{M. {Portail} {et~al.}(2015){Portail}, {Wegg}, {Gerhard}, \&
  {Martinez-Valpuesta}}]{portai_etal_2015}
{Portail}, M., {Wegg}, C., {Gerhard}, O., \& {Martinez-Valpuesta}, I. 2015,
  \bibinfo{title}{{Made-to-measure models of the Galactic box/peanut bulge:
  stellar and total mass in the bulge region},} \mnras, 448, 713,
  \dodoi{10.1093/mnras/stv058}

\bibitem[{N. {Raha} {et~al.}(1991){Raha}, {Sellwood}, {James}, \&
  {Kahn}}]{raha_etal_1991}
{Raha}, N., {Sellwood}, J.~A., {James}, R.~A., \& {Kahn}, F.~D. 1991,
  \bibinfo{title}{{A dynamical instability of bars in disk galaxies},} \nat,
  352, 411, \dodoi{10.1038/352411a0}

\bibitem[{A.~B. {Romeo} {et~al.}(2023){Romeo}, {Agertz}, \&
  {Renaud}}]{romeo_etal_2023}
{Romeo}, A.~B., {Agertz}, O., \& {Renaud}, F. 2023, \bibinfo{title}{{The
  specific angular momentum of disc galaxies and its connection with galaxy
  morphology, bar structure, and disc gravitational instability},} \mnras, 518,
  1002, \dodoi{10.1093/mnras/stac3074}

\bibitem[{K. {Saha} \& T. {Naab}(2013){Saha} \& {Naab}}]{sah_naa_2013}
{Saha}, K., \& {Naab}, T. 2013, \bibinfo{title}{{Spinning dark matter haloes
  promote bar formation},} \mnras, 434, 1287, \dodoi{10.1093/mnras/stt1088}

\bibitem[{A. {Saintonge} \& B. {Catinella}(2022){Saintonge} \&
  {Catinella}}]{sai_cat_2022}
{Saintonge}, A., \& {Catinella}, B. 2022, \bibinfo{title}{{The Cold
  Interstellar Medium of Galaxies in the Local Universe},} \araa, 60, 319,
  \dodoi{10.1146/annurev-astro-021022-043545}

\bibitem[{B.~M. {Sch{\"a}fer}(2009){Sch{\"a}fer}}]{sch_bjo_2009}
{Sch{\"a}fer}, B.~M. 2009, \bibinfo{title}{{Galactic Angular Momenta and
  Angular Momentum Correlations in the Cosmological Large-Scale Structure},}
  International Journal of Modern Physics D, 18, 173,
  \dodoi{10.1142/S0218271809014388}

\bibitem[{D.~W. {Sciama}(1955){Sciama}}]{sciama_etal_1955}
{Sciama}, D.~W. 1955, \bibinfo{title}{{On the formation of galaxies in a steady
  state universe},} \mnras, 115, 3, \dodoi{10.1093/mnras/115.1.3}

\bibitem[{J.~A. {Sellwood} \& O. {Gerhard}(2020){Sellwood} \&
  {Gerhard}}]{sel_ger_2020}
{Sellwood}, J.~A., \& {Gerhard}, O. 2020, \bibinfo{title}{{Three mechanisms for
  bar thickening},} \mnras, 495, 3175, \dodoi{10.1093/mnras/staa1336}

\bibitem[{J. {Shen} {et~al.}(2010){Shen}, {Rich}, {Kormendy}, {Howard}, {De
  Propris}, \& {Kunder}}]{shen_etal_2010}
{Shen}, J., {Rich}, R.~M., {Kormendy}, J., {et~al.} 2010, \bibinfo{title}{{Our
  Milky Way as a Pure-disk Galaxy{\textemdash}A Challenge for Galaxy
  Formation},} \apjl, 720, L72, \dodoi{10.1088/2041-8205/720/1/L72}

\bibitem[{K. {Sheth} {et~al.}(2012){Sheth}, {Melbourne}, {Elmegreen},
  {Elmegreen}, {Athanassoula}, {Abraham}, \& {Weiner}}]{sheth_etal_2012}
{Sheth}, K., {Melbourne}, J., {Elmegreen}, D.~M., {et~al.} 2012,
  \bibinfo{title}{{Hot Disks and Delayed Bar Formation},} \apj, 758, 136,
  \dodoi{10.1088/0004-637X/758/2/136}

\bibitem[{L.~A. {Silva-Lima} {et~al.}(2022){Silva-Lima}, {Martins}, {Coelho},
  \& {Gadotti}}]{lima_etal_2022}
{Silva-Lima}, L.~A., {Martins}, L.~P., {Coelho}, P. R.~T., \& {Gadotti}, D.~A.
  2022, \bibinfo{title}{{Revisiting the role of bars in AGN fuelling with
  propensity score sample matching},} \aap, 661, A105,
  \dodoi{10.1051/0004-6361/202142432}

\bibitem[{B.~D. {Simmons} {et~al.}(2014){Simmons}, {Melvin}, {Lintott},
  {Masters}, {Willett}, {Keel}, {Smethurst}, {Cheung}, {Nichol}, {Schawinski},
  {Rutkowski}, {Kartaltepe}, {Bell}, {Casteels}, {Conselice}, {Almaini},
  {Ferguson}, {Fortson}, {Hartley}, {Kocevski}, {Koekemoer}, {McIntosh},
  {Mortlock}, {Newman}, {Ownsworth}, {Bamford}, {Dahlen}, {Faber},
  {Finkelstein}, {Fontana}, {Galametz}, {Grogin}, {Gr{\"u}tzbauch}, {Guo},
  {H{\"a}u{\ss}ler}, {Jek}, {Kaviraj}, {Lucas}, {Peth}, {Salvato}, {Wiklind},
  \& {Wuyts}}]{simmon_etal_2014}
{Simmons}, B.~D., {Melvin}, T., {Lintott}, C., {et~al.} 2014,
  \bibinfo{title}{{Galaxy Zoo: CANDELS barred discs and bar fractions},}
  \mnras, 445, 3466, \dodoi{10.1093/mnras/stu1817}

\bibitem[{V. {Springel} {et~al.}(2021){Springel}, {Pakmor}, {Zier}, \&
  {Reinecke}}]{spring_etal_2021}
{Springel}, V., {Pakmor}, R., {Zier}, O., \& {Reinecke}, M. 2021,
  \bibinfo{title}{{Simulating cosmic structure formation with the GADGET-4
  code},} \mnras, 506, 2871, \dodoi{10.1093/mnras/stab1855}

\bibitem[{E. {Vasiliev}(2019){Vasiliev}}]{vasili_2019}
{Vasiliev}, E. 2019, \bibinfo{title}{{AGAMA: action-based galaxy modelling
  architecture},} \mnras, 482, 1525, \dodoi{10.1093/mnras/sty2672}

\bibitem[{M. {Vitvitska} {et~al.}(2002){Vitvitska}, {Klypin}, {Kravtsov},
  {Wechsler}, {Primack}, \& {Bullock}}]{vitvit_etal_2002}
{Vitvitska}, M., {Klypin}, A.~A., {Kravtsov}, A.~V., {et~al.} 2002,
  \bibinfo{title}{{The Origin of Angular Momentum in Dark Matter Halos},} \apj,
  581, 799, \dodoi{10.1086/344361}

\bibitem[{S.~D.~M. {White}(1984){White}}]{white_1984}
{White}, S.~D.~M. 1984, \bibinfo{title}{{Angular momentum growth in
  protogalaxies},} \apj, 286, 38, \dodoi{10.1086/162573}

\bibitem[{L.~M. {Widrow} {et~al.}(2008){Widrow}, {Pym}, \&
  {Dubinski}}]{widrow_etal_2008}
{Widrow}, L.~M., {Pym}, B., \& {Dubinski}, J. 2008, \bibinfo{title}{{Dynamical
  Blueprints for Galaxies},} \apj, 679, 1239, \dodoi{10.1086/587636}

\bibitem[{T. {Worrakitpoonpon}(2025){Worrakitpoonpon}}]{worrak_2025}
{Worrakitpoonpon}, T. 2025, \bibinfo{title}{{Bar Instability and Formation
  Timescale across Toomre's Q Parameter and Central Mass Concentration: Slow
  Bar Formation or True Stability},} \apj, 979, 166,
  \dodoi{10.3847/1538-4357/ada35f}

\bibitem[{J. {Zjupa} \& V. {Springel}(2017){Zjupa} \&
  {Springel}}]{zju_spr_2017}
{Zjupa}, J., \& {Springel}, V. 2017, \bibinfo{title}{{Angular momentum
  properties of haloes and their baryon content in the Illustris simulation},}
  \mnras, 466, 1625, \dodoi{10.1093/mnras/stw2945}

\end{thebibliography}
\bibliographystyle{aasjournalv7}

\end{document}